\documentclass[aps,prl,showpacs,twocolumn,reprint,floats,epsfig,pdflatex]{revtex4-1}
\usepackage{epsfig}
\usepackage{amsmath}
\usepackage{amsfonts}
\usepackage{graphicx}
\usepackage{amssymb}
\usepackage{amsbsy}
\usepackage{subfigure}
\usepackage{tabularx}

\newcommand{\bra}[1]{\langle{#1}|}
\newcommand{\ket}[1]{|{#1}\rangle}
\newcommand{\braket}[1]{\langle{#1}\rangle}

\begin{document}

\title {Precision measurement of branching fractions of $^{138}$Ba$^{+}$: Testing many body theories below one percent level}

\author{D.~De~Munshi$^1$, T.~Dutta$^1$, R.~Rebhi$^1$ and M.~Mukherjee$^{1,2*}$}
\affiliation{$^1$Centre for Quantum Technologies, National
University Singapore, Singapore 117543.\\
$^2$ Department of Physics, National University Singapore, Singapore 117551. }

\date{\today}

\begin{abstract}
The branching fractions from the excited state $6P_{1/2}$ of singly charged barium ion has been measured with a precision $0.05\%$ in an ion trap experiment. This measurement along with the known value of the upper state life-time allowed the determination of the dipole matrix elements for the transitions $P-S$ and $P-D$ to below one percent level. Therefore, for the first time it is now possible to compare the many body calculations of these matrix elements at level which is of significance to any parity non-conservation experiment on barium ion. Moreover, these dipole matrix elements are the most significant contributors to the parity violating matrix element between the $S-D$ transition, contributing upto $90\%$ to the total. Our results on the dipole matrix elements are $3.306\pm0.014$ and $3.036\pm0.016$ for the $S-P$ and $P-D$ transitions respectively.

\end{abstract}


\maketitle

Trapping and laser cooling of ions provide perturbation free environment to measure atomic state life-time~\citep{Olm09}, light-shift~\citep{She08}, branching ratio~\cite{Kur08} and other fundamental properties of atoms with high precision~\cite{Roo06}. This leads to the use of trapped and laser cooled ions for quantum state manipulation~\cite{Nig14,Wil14}, atomic clocks~\cite{Cho10} and to study fundamental interactions~\cite{Lea11}. The study of fundamental interactions via atomic properties include measurements of the Lamb shift~\cite{Lev72}, the parity non-conservation~(PNC) in atomic system~\cite{Woo97}, the conserved vector current hypothesis~\citep{Man04}, the electron-electric dipole moment (e-EDM)~\cite{Hud11} {\it etc}.. As most of the original experiments have been carried out with atomic beams, they suffered from large systematic uncertainties due to limited control over the environment. These systematics are largely absent for stored atomic systems, in addition, they provide long observation time. Therefore, in recent years, trapped and laser cooled ions have emerged as potential candidate to perform high precision experiments of fundamental importance like atomic parity violation~\cite{For93,Sah11} and e-EDM~\cite{Lea11}. Barium ion is particularly suitable for the investigation of PNC as was pointed out by Fortson~\cite{For93} because of its large nuclear charge and ease of laser cooling and trapping. 

The best atomic PNC measurement performed so far is that of cesium with a precision of 0.3\%~\cite{Woo97}. However, the nuclear anapole moment obtained from this measurement shows a discrepancy with other nuclear data strongly suggesting the need to measure atomic PNC in other species in order to verify or to go beyond the Standard Model of particle physics. In this context, a number of experimental groups are pursuing an ion trap based atomic PNC experiment which has been proposed to be capable of limiting systematic uncertainty to below one percent level. In addition to the experiment, one also needs the theoretical value of the parity violating dipole matrix element with a similar precision. In principle different variants of the coupled cluster theory~\cite{Gue91,Dzu01,Gee02,Sah07} are capable of providing such precision, provided the many-body wavefunctions are accurately known. Precision measurement of atomic properties of the low lying energy levels allow these theory to be tuned to provide high accuracy wavefunctions. Therefore measuring branching fractions or transition probabilities and life-time of Ba$^+$ with precision below one percent are of utmost need. Moreover, for PNC measurement in barium ion between the states $\ket{6S_{\frac{1}{2}}}$ and $\ket{5D_{\frac{3}{2}}}$, the contribution to the parity non-conserving dipole matrix element $\epsilon^{PNC}$ comes from the sum over all high $p-$states given by, 

\begin{equation}
\epsilon^{PNC}=\sum_{n,j}\frac{\bra{5D_{3/2}}\hat{d}\ket{nP_{j}}\bra{nP_{j}}H^{PNC}\ket{6S_{1/2}}}{W_{6S_{1/2}}-W_{nP_{j}}}+h.c.,
\label{eq3}
\end{equation}

where, $\hat{d}$, $H^{PNC}$ and $W$ are the dipole operator, PNC operator and electronic binding energy respectively. The principle quantum number and the total angular momentum quantum numbers are denoted by $n$ and $j$. The state $\ket{6P_{\frac{1}{2}}}$ contributes about 90\%~\cite{Dzu01,Sah07} to the PNC matrix element $\epsilon^{PNC}$ via the matrix elements $\bra{6P_{\frac{1}{2}}}D\ket{5D_{\frac{3}{2}}}$ and $\bra{6P_{\frac{1}{2}}}D\ket{6S_{\frac{1}{2}}}$ as shown in eq.~(\ref{eq3}).\\

In this letter, we present measurements of the branching fractions for the dipole transitions from the $6P_{\frac{1}{2}}$ state of barium ion with a precision well below $0.1\%$ thereby making it possible to compare with the existing theory to the precision that is required for any PNC measurement with barium ion. The branching fraction and transition probabilities of a fast decaying excited state can be measured by different techniques like ultra-fast excitations~\cite{Kur08}, optical nutation~\cite{Kas93} or by simple photon counting at different wavelengths. The first approach requires complicated laser pulse sequence and suffers from systematics due to synchronization, the second one is prone to systematics due to the measurement of the actual intensity at the ion position and the last technique is limited by the availability of well calibrated detectors. We performed the branching fraction measurement on barium ion using a similar method as recently proposed and performed for calcium by Ramm {\it et al.}~\cite{Ram13}. This method has been shown to be largely free of common systematics like magnetic field fluctuation, intensity fluctuations {\it etc.}. In the following, a brief description of the experimental setup, followed by procedure, results and discussion will be made. 

\begin{figure}
\includegraphics[width=\linewidth]{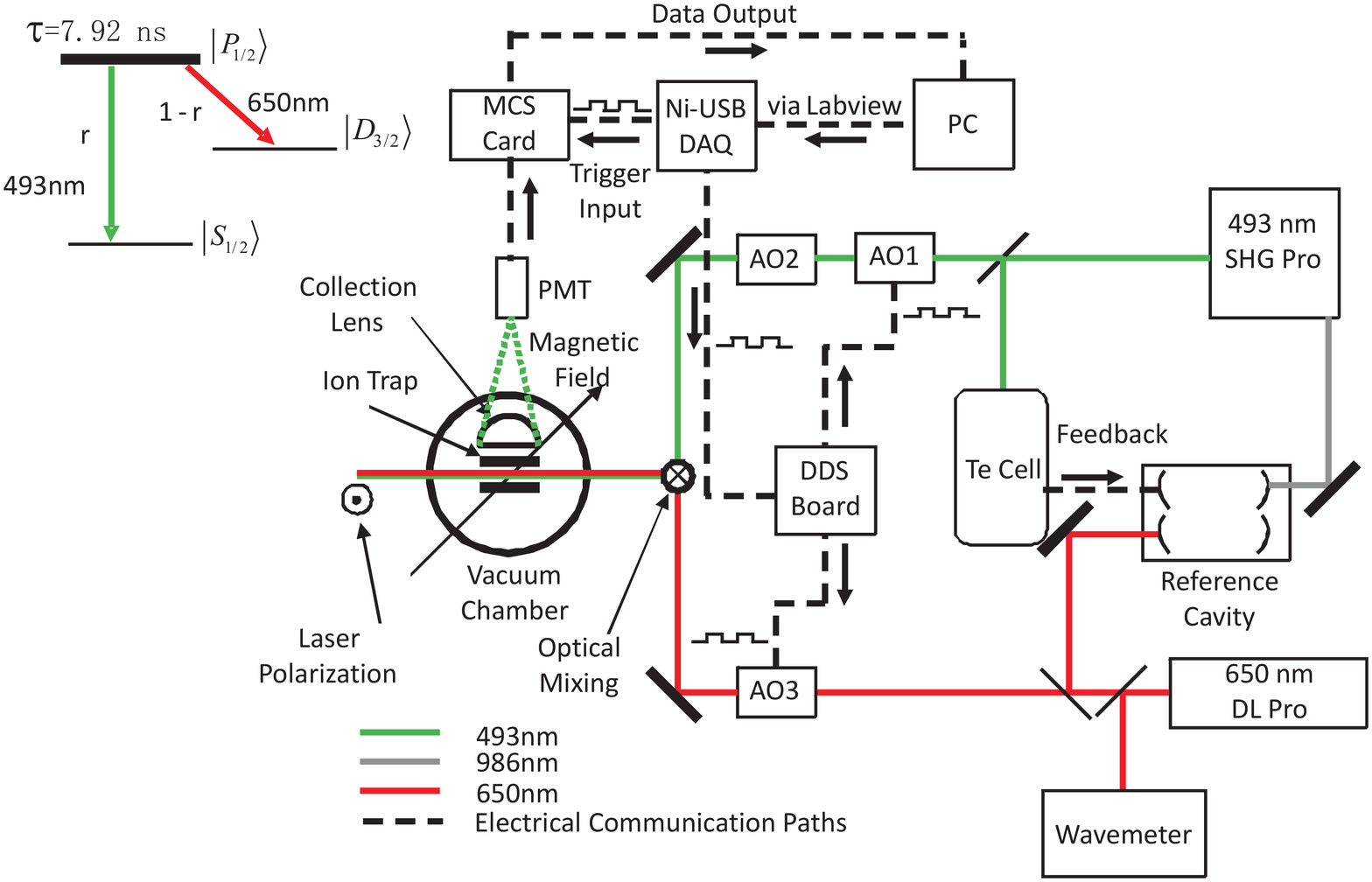}
\caption{(Color online) Schematic diagram of our experiment: During the experiment, AO1 and AO3 are switched using a DDS controlled by NIDAQ. Relevant atomic levels of Ba$^+$ ions are also shown. The branching fractions reported here are $r$ and $1-r$ for the two decay channels while the upper state life-time is the best literature value available~\cite{Kus78}.} \label{Fig1}
\end{figure}

A schematic of our setup is shown in Figure(\ref{Fig1}). The ion trap is a linear blade trap with radial parameter $r_0=0.7$~mm and axial parameter $z_0=2.2$~mm. The trap is operated at a radio-frequency of $16$~MHz providing ions' motional secular frequencies of about $1$~MHz in the radial direction while several hundred kHz in the axial direction. Although the experimental results presented here are independent of the number and Coulomb crystal structure of the ions, we have used a linear chain of ions ranging from a few to about 20 ions. The cooling laser is a frequency doubled diode laser from {\it Toptica SHGpro} providing light at $493$~nm (green) which addresses the main cooling transition between $S_{\frac{1}{2}}$ and $P_{\frac{1}{2}}$. In order to minimize laser frequency drifts during the whole experiment, the laser frequency is locked to a reference cavity which is then locked to one of the closest molecular transitions of $^{130}$Te$_{2}$~\cite{Raa98} by modulation transfer spectroscopy. The frequency difference between resonance of $^{130}$Te$_{2}$ and $^{138}$Ba$^{+}$ is bridged by an acusto-optic modulator (AOM) named AO1 in double-pass configuration and another AOM (AO2) in single-pass. As the cooling transition is not entirely closed due to the presence of a metastable $D_{\frac{3}{2}}$ state, the ion population is re-pumped into the cooling cycle by a $650$~nm (red) {\it Toptica DLpro} laser. The $650$ nm laser is locked to a reference cavity which has a common zerodur$^\copyright$ spacer with the cavity of $493$ nm, thereby cancelling their relative drifts. In order to avoid population trapping into the Zeeman dark states, a magnetic field of about $1.7$~G is applied by external coils. As shown in figure~(\ref{Fig1}), spontaneously emitted photons are collected perpendicular to the cooling beam. Barium ions are created by two step photo-ionization process consisting of a resonant excitation to an inter-combination line of neutral barium and then to continuum by a home-built external cavity diode laser at $413$ nm. In order to maximize the photon counts, we collect the fluorescence photons from the ions using an in-vacuum large numerical aperture (NA$=0.4$) aspheric lens from {\it Asphericon}. The spontaneously emitted photons are counted by a Hamamatsu photo multiplier tube~(PMT) after being filtered by an interference filter at $493$~nm with a bandwidth of $20$~nm from $Semrock$. 


If we consider the probability of spontaneous emission of photons from the state $\ket{6P_{1/2}}$ to $\ket{6S_{1/2}}$ as $r$, the probability of emission into the state $\ket{5D_{3/2}}$ is $1-r$. Now if only the green beam is on, there will be on average $\braket{n}$ number of photons emitted before the ion settles to the state $\ket{5D_{3/2}}$. Therefore the number of average green photons emitted is ~\cite{Ram13}

\begin{equation}\label{eq1}
\braket{N_g}=\braket{n}-1=\frac{r}{1-r}.
\end{equation}

\begin{figure}
\includegraphics[width=\linewidth]{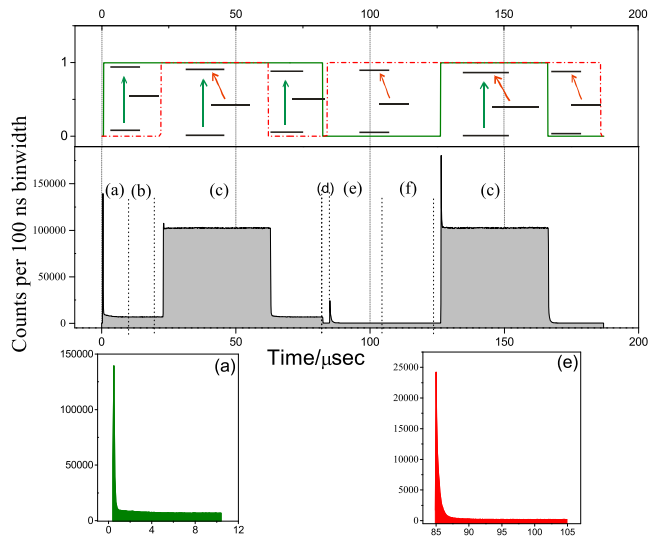}
\caption{(Color online)Experimental sequence and green photon count measurement. The top panel shows the experimental time sequence of the green and red pulses while the corresponding photon counts are shown in the middle panel. The sequence consists of (a) green photon measurement (green laser on), (b) green background counts (green laser on), (c) cooling (both lasers on), (d) optical pumping to $D$-state (green laser on), (e) red-repumping while green photon counting (red laser on), (f) dark count measurement (both lasers off) and (c) another cooling pulse. The lower panel shows zoomed part of both the decay curves due to transfer and back-transfer of population.} \label{Fig3}
\end{figure}

Once we have transferred the ion with only the green laser beam, we can perform a similar back transfer from the $\ket{5D_{3/2}}$ to the $\ket{6P_{1/2}}$ state by applying the red laser beam alone. During this transfer and back transfer, we measure the average number of green photons emitted. Thus we get 

\begin{equation}\label{eq2}
r=\frac{N_g}{N_g+N_r},
\end{equation}

where $r$ is the branching fraction for the $6P_{1/2}-6S_{1/2}$ transition and $N_r$ is the total counts of green photons while the red laser is kept on. As is evident from eq.(\ref{eq1}) and eq.(\ref{eq2}), there is no dependence of $r$ on the intensity and detuning of the excitation laser or the efficiency of the detector. The time sequence used to implement the scheme is shown in figure~(\ref{Fig3}). The average count $N_g$ and $N_r$ are measured for $10~\mu$s and $20~\mu$s in steps (a) and (e) respectively. An equivalent time is also spent to collect background counts for subtraction to obtain the actual counts of $N_g$ and $N_r$ originating from the ions. 
  The time sequence is controlled by National Instrument DAQ (NIDAQ/USB 6363), while the photon counts are registered into a multi channel scalar~(MCS) from {\it Ortec} with a resolution of $100$~ns. To work in the linear region of the detector's response, the intensity of the green laser has been kept low throughout these measurements. The pulsing of the lasers is done by intensity modulating the AO1 for the green and AO3 for the red using a direct digital synthesizer (DDS) which drives the AOMs via amplifiers. \\
 We repeated the experimental sequence as mentioned in figure~(\ref{Fig3}) for $27$ different experimental sets, each of these measurements have $2 \times 10^6$ cycles. The total time required is mainly limited by the required uncertainty which we targeted to be below $0.1\%$ for the branching fraction measurement. In order to check for systematics we have performed about $50$ similar experiments under different experimental conditions like varying magnetic field, laser intensity, added micro-motion, different Coulomb crystal structure {\it etc.}. None of the above varied condition showed variation in the value of $r$ above the $1\sigma$ statistical variation. The laser intensity, however needs to be within certain range as too high value would paralyze the PMT in short time scales and too low value would make the decay exponent too long for counting in reasonable time. In order to check for any birefringence in our detection setup leading to disproportionate red to green counts due to any possible polarization dependence, we purposely changed the polarization angle of the linearly polarized green and the red beams showing no significant deviation beyond the statistical uncertainty. Even though the PMT is in the non-paralyzing regime, a significant contribution to our systematic uncertainty comes from the detector dead time. The dead time of the PMT leads to a calculable shift as well as uncertainty. The systematic shifts and uncertainties are tabulated in table(\ref{tab1}). As is evident, the main limitation to the measured uncertainty is statistical and hence, limited by the finite measurement time. A relatively large shift of $r$ is due to the detector dead-time of $70~$ns combined with lower resolution in our MCS binning (about $100~$ns). Nevertheless, the uncertainties are similar to the statistical uncertainties thereby not limiting the final values.

\begin{table}
  \caption{Error budget for the $S-P$ branching fraction measurement.}
  \begin{tabularx}{0.5\textwidth}{lXX}
    \hline\hline
    {\bf parameter} & {\bf shift} & {\bf uncertainty} \\
    \hline
    Detector dead time $70~ns$\footnote{Hamamatsu H7421-40}\quad  & $+5\times10^{-3}$ & $9\times10^{-4}$ \\\\
    Photon counting (statistical)			& & $1.9\times10^{-4}$ \\\\
    Photon counting \\(finite measurement time)	&$6.6\times10^{-6}$ & $3.3\times10^{-6}$  \\
\hline\hline
  \end{tabularx}
  \label{tab1}
\end{table}

\begin{table*}
  \caption{A comparison of the values of the branching fractions, transition probabilities and matrix elements for Ba$^+$ between different experiments and theories arranged chronologically.}
  \begin{tabularx}{\textwidth}{XXXXX}
    \hline\hline
    {\bf Transition involved} & {\bf Branching fraction} & {\bf Transition probability} $\times 10^8 s^{-1}$ & {\bf Transition matrix (a.u.)} & {\bf reference}\\
    \hline
    $\ket{P_{\frac{1}{2}}}-\ket{S_{\frac{1}{2}}}$ 	& $0.735\pm0.021$ & $0.95\pm 0.07$ & $3.35\pm 0.11$&\cite{Gal67} \\
        											& & $0.955\pm 0.095$ & $3.36\pm 0.15$ & \cite{Rea80} \\
        											& & $0.9178$ & $3.300$ &\cite{Gue91}  \\
        											& $0.756\pm0.012$ & $0.953\pm 0.024$ & $3.362\pm 0.038$ &\cite{Dav92}  \\
        											& & $0.95\pm 0.09$ & $3.36\pm 0.14$ &\cite{Kas93}  \\
													& & $0.9232$ & $3.309$ &\cite{Dzu01}  \\
													& & $0.9368$ & $3.333$ &\cite{Gee02}  \\
													& & $0.953\pm 0.095$ & $3.36\pm 0.15$ &\cite{Cur04}  \\  
													& & $0.978$ & $3.405$ &\cite{Sah07}  \\
													& $0.7304\pm0.0004$ & $0.922\pm 0.009$ & $3.306\pm 0.014$ & this work \\      										        											\hline
    $\ket{P_{\frac{1}{2}}}-\ket{D_{\frac{3}{2}}}$ 	& $0.265\pm0.021$ & $0.33\pm 0.04$ & $2.99\pm 0.18$&\cite{Gal67} \\
        											& & $0.332\pm 0.083$ & $3.00\pm 0.37$ & \cite{Rea80} \\
        											& & $0.334$ & $3.007$ & \cite{Gue91}  \\
        											&$0.244\pm0.012$ & $0.3097\pm 0.018$ & $2.895\pm 0.084$ & \cite{Dav92}  \\
        											& & $0.338\pm 0.019$ & $3.025\pm 0.085$ & \cite{Kas93}  \\
													& & $0.37$ & $3.165$ & \cite{Dzu01}  \\
													& & $0.326$ & $2.971$ & \cite{Gee02}  \\
													& & $0.31\pm 0.031$ & $2.90\pm 0.15$ &\cite{Cur04}  \\  
													& & $0.331$ & $2.993$ & \cite{Sah07}  \\
													& $0.2696\pm0.0004$ & $0.3404\pm 0.0035$ & $3.036\pm 0.016$ & this work \\     
    											 	\hline\hline
  \end{tabularx}
  \label{tab2}
\end{table*}

\begin{figure}
\includegraphics[width=1.0\linewidth]{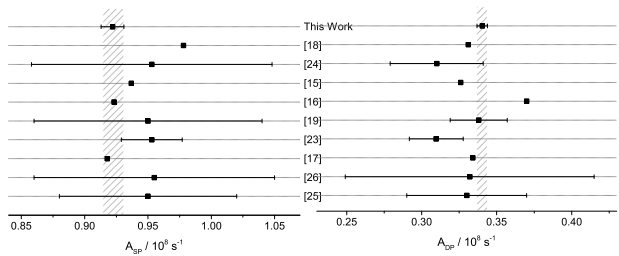}
\caption{A comparison of different measurements and theory values of the dipole transition probabilities for $S-P$ and $D-P$ transitions. The hatched area provides the value and $1\sigma$ confidence band of this measurement. The values which donot have error associated with them are theory values while values by \cite{Cur04,Rea80} are compilation results of earlier experiments.} \label{Fig4}
\end{figure}

The measured values of the branching fractions, along with the best literature value of the upper state life-time, $7.92\pm0.08$~ns~\cite{Kus78}, provides the transition probability as well as the matrix elements of the relevant transitions by following the procedure in~\cite{Gee02}. In table(\ref{tab2}), we show our results along with the values measured or calculated till date. The best measured experimental data on these transitions are limited to about $5\%$ uncertainties on the matrix elements. On the contrary, since the experimental proposal by Fortson~\cite{For93} for the possibility of measuring the atomic PNC in barium ion, the accuracies on the theoretical values of the matrix elements has improved significantly aiming towards below one percent level where many-electron correlation effects become significant. As can be seen from figure~(\ref{Fig4}), the theory values scatter within the experimental uncertainties while the claimed theoretical uncertainties are significantly lower than experiments~\cite{Sah07}. Our measurements, for the first time, provides the values below one percent limit, thereby allowing the theories to be compared at a similar uncertainty level. The $S-P$ transition probability is rather close to the theory values of~\cite{Dzu01,Gue91,Gee02} but it is quite off from the value of~\cite{Sah07}. On the other hand, the $P-D$ transition probability is close to the values calculated by~\cite{Gue91,Gee02,Sah07} while deviating significantly from \cite{Dzu01}. These theories mostly consider all orders in perturbation but limited to certain number of collective excitations, therefore it is now possible to make a comparative study of these different approaches in view of the experimental data. The branching fractions themselves are important for estimating the abundance of barium in solar and stellar atmosphere~\cite{Dav92} which provides insight into the process of nucleo-synthesis, especially of heavy elements. Our measured branching fractions are $0.7304\pm0.0004$ and $0.2696\pm0.0004$ for the $P-S$ and $P-D$ branches respectively.   

In conclusion, we measured the matrix elements that contribute the maximum ($90\%$) to the atomic parity violation mixing for the otherwise forbidden $S-D$ transition in ionic barium. As it has been pointed out in \cite{Sah07}, so far high precision theoretical values could not be compared to previous experiment due to relatively large uncertainties in the experimental results. However, it is now possible to clearly distinguish between different theories as is evident from figure~(\ref{Fig4}). Although our results match well with all the previous experimental results, the precision is $6$ and $2.5$ times better than the previous best measurement. Moreover, our result differs significantly from the theory value of \cite{Sah07} where the precision is quoted to be below one percent for the $S-P$ matrix element while the other theory results do not specify their uncertainties. The overall uncertainty in the matrix element determination is limited by the precision of the life-time of the upper state ($\sim 1\%$), therefore it is in principle possible to improve the uncertainty further by performing more precise measurements of the life-time. Our branching fraction measurement is precise to $0.05\%$ which is a parameter often used to quantify the abundance of barium in solar and stellar atmospheres~\cite{Dav92}. Therefore, the branching ratio measurements will contribute to the better understanding of element formation in stars while the precise measurement of the matrix elements will contribute towards the verification of the Standard Model or to go beyond it using atomic PNC as a tool. 

We acknowledge the support by the National Research Foundation and the Ministry of Education, Singapore to carry out this work.

\end{document}